\documentclass{acm_proc_article-sp}
\pdfoutput=1

\usepackage{moreverb}

\usepackage{amssymb}
\usepackage{subfigure}
\usepackage{color}
\usepackage{listings}
\usepackage[utf8x]{inputenc}
\usepackage{multirow}
\usepackage{array}
\usepackage{float}
\usepackage{url}
\usepackage{amsmath}

\usepackage[dvips,colorlinks,bookmarksopen,bookmarksnumbered,citecolor=red,urlcolor=red]{hyperref}

\begin{document}

\title{Elliptic Curve Based Zero Knowledge Proofs and Their Applicability on Resource Constrained Devices \thanks{This work has been partially supported by the European Union under contract number ICT-2010-258885 (\textsf{SPITFIRE}).}}

\author{
\alignauthor
Ioannis Chatzigiannakis, Apostolos Pyrgelis, Paul G. Spirakis, Yannis C. Stamatiou \\
       \affaddr{Research Academic Computer Technology Institute}\\
       \affaddr{and Computer Engineering and Informatics Department, University of Patras}\\
       \affaddr{Greece}\\
       \email{\{ichatz,pyrgelis,spirakis,stamatiou\}@cti.gr}
}

\maketitle

\begin{abstract}
Elliptic Curve Cryptography (ECC) is an attractive alternative to
conventional public key cryptography, such as RSA. ECC is an ideal 
candidate for implementation on constrained devices where the major 
computational resources i.e. speed, memory are limited and
low-power wireless communication protocols are employed.
That is because it attains the same security levels with traditional 
cryptosystems using smaller parameter sizes. Moreover, in several 
application areas such as person identification and eVoting, it is 
frequently required of entities to prove knowledge of some fact without 
revealing this knowledge. Such proofs of knowledge are called Zero Knowledge 
Interactive Proofs (ZKIP) and involve interactions between two communicating 
parties, the Prover and the Verifier. In a ZKIP, the Prover demonstrates the possesion 
of some information (e.g. authentication information) to the Verifier without disclosing it. 
In this paper, we focus on the application of ZKIP protocols on resource 
constrained devices. We study well-established ZKIP protocols based on the 
discrete logarithm problem and we transform them under the ECC setting. 
Then, we implement the proposed protocols on Wiselib, a generic and open source 
algorithmic library. Finally, we present a thorough evaluation of the protocols 
on two popular hardware platforms equipped with low end microcontrollers 
(Jennic JN5139, TI MSP430) and 802.15.4 RF transceivers, in terms of code size, execution 
time, message size and energy requirements. To the best of our knowledge, this 
is the first attempt of implementing and evaluating ZKIP protocols with emphasis on 
low-end devices. This work's results can be used from developers who wish to achieve
certain levels of security and privacy in their applications.
\end{abstract}


\section{Introduction}

Recent advances in wireless communications and microelectro systems
have lead to the construction of tiny devices with strong processing
and communication capabilities. Mobile phones, PDAs, sensor devices 
and RFID tags are becoming a part of our daily lives and their
networked interconnection makes the vision of the Internet of Things
a real situation. Smart towns and buildings, automated hospitals,
intelligent cars and sensors embedded in clothes are concepts that
are already being developed.

However, the wireless nature of communication that these devices provide
(e.g. 802.11, 802.15.4, Bluetooth) and the fact that there is not a fixed 
infrastructure in such dynamic networks raise significant security 
and trust issues \cite{Chen_asurvey}. In some cases, these petit computers 
may need to exchange crucial information that needs to be protected.
Adversaries equipped with strong computers and antennae can eavesdrop, analyze 
or alter the data exchanged. Terms like information security, data confidentiality 
and integrity, entity authentication and identification need to be considered regarding
the wireless setting \cite{4625802}. The field of cryptography offers some solutions 
to the above issues but they need to be adapted suitably for their application on embedded devices.

As the model of ubiquitous computing arises, the use of low-end devices on a daily basis
increases (e.g. access to a social network from a mobile phone). Thus, users will need a way to
preserve their privacy and not reveal information that could be exploited by adversaries. 
For such issues, cryptography offers the tool of zero-knowledge proofs. A zero-knowledge 
proof can be used whenever someone needs to prove the possession of critical data without 
exchanging or revealing the actual data. Examples of today's Internet applications that 
use zero-knowledge proofs are e-commerce, e-voting, access authorization and entity authentication. 

There exist various kinds of zero-knowledge proofs that involve problems like
graph isomorphism and integer factorization. In this paper, we focus on the study of 
well-established zero-knowledge protocols based on the discrete logarithm problem. 
Up to now, although a wide variety of zero-knowledge protocols of this category has been proposed, 
( e.g. see \cite{Smith05cryptographymeets} ) no actual implementations regarding resource constrained 
devices have been presented. For this reason and regarding authentication and privacy issues on the Internet of Things, 
we concentrate on the application of zero-knowledge protocols for the security and privacy empowerment of
current wireless networks consisting of low constrained devices.

Nevertheless, when considering low-end devices capable of communicating wirelessly one should take into
account the resource limitations, i.e. the restricted processing power and memory as well as the 
particularities imposed by the low-power wireless communication protocols (e.g. packet loss, channel 
throughput, message size). One can realize that high computation as well as high communication overhead 
leads to great energy consumption which is another very important constraint. These limitations consist the actual 
challenge, when trying to implement heavy protocols in terms of computation and communication, like zero-knowledge proofs, 
on constrained devices. Thus, in this work we emphasize on the elliptic curve cryptography approach, 
as proposed in \cite{DBLP:conf/wecwis/AlmuhammadiSM04}, with the implementation of zero-knowledge protocols on 
constrained devices as our main objective.

\subsection{Our Contributions}

The contributions of this work are threefold.

Firstly, we study well-established zero-knowledge protocols based on the discrete logarithm
problem (DLP) and we show how these protocols can be transformed and
adapted under the elliptic curve discrete logarithm problem (ECDLP)
and we state why these transformations are correct. This transformation step required a 
careful examination of the logarithmic exponential 
operations and the corresponding scalar multiplication on the elliptic curves as well as
the arithmetic operations on the respective number fields.
Such an adaptation consists the key for implementing zero-knowledge protocols on embedded devices. 
That is because elliptic curve cryptography (ECC) offers the same level of security with
other public key cryptosystems (e.g. RSA) with the use of much
smaller keys (see \textbf{Appendix}). This advantage ensures that we save space on the
limited memory of such tiny devices, that the protocols' compiled
code can actually fit well on them and that the protocols' message sizes are reasonable. 

Secondly, we implement the new proposed protocols on Wiselib, a generic and open source algorithmic library.
This way our code is generic, highly portable, publically available and ready to be used by developers that 
wish to provide certain levels of security and privacy in their applications. 

Finally, we present a thorough evaluation of the new zero-knowledge protocols on two popular hardware platforms
equipped with widely used low-end microcontrollers (Jennic JN5139, TI MSP430) as well as 802.15.4 \cite{802154}
RF transceivers in terms of execution time, code size, messages' size and energy consumption. 
We experimentally prove that our protocols have small code footprint (around 8Kb) and that their 
exchanged messages fit well in the technical specifications of the 802.15.4 protocol.
To the best of our knowledge, this is the first attempt of implementing and
evaluating zero-knowledge protocols, with emphasis on low-constrained devices. 
Moreover, the resulting library can form the basis for the implementation of more complex 
protocols employed in various cryptographic applications, like attribute based credentials 
\cite{Brands:2000:RPK:517876}. 

\subsection{Paper Outline}

The remaining of our paper is structured as follows: Firstly, in \textbf{Section 2} 
we present an overview of zero-knowledge proofs and in \textbf{Section 3} 
we show how well established zero knowledge protocols based on the discret logarithm problem 
can be adapted on the elliptic curve discrete logarithm problem. In \textbf{Section
4} we refer to the Wiselib platform on which we implemented the
protocols, the hardware used for our experiments and we present an
evaluation of the protocols on actual constrained devices. 
In \textbf{Section 5} we describe three everyday Internet of Things applications where zero-
knowledge proofs can be used. In \textbf{Section 6} we conclude 
and propose some of our ideas for future work. Finally, for a reader's information
in the \textbf{Appendix} we refer to the basic definitions of elliptic curve cryptography 
and the reasons that make it suitable for constrained devices.
\section{An Overview of Zero Knowledge Protocols}

Generally, a zero-knowledge protocol allows a proof of the truth of
an assertion, while conveying no information whatsoever about the
assertion itself other than its actual truth \cite{548089}. Usually,
such a protocol involves two entities, a prover and a verifier. A
zero-knowledge proof allows the prover to demonstrate knowledge of a
secret while revealing no information whatsoever of use to the
verifier in conveying this demonstration of knowledge to others.

The zero-knowledge protocols to be discussed are instances of
interactive proof systems and non-interactive proof systems. In the
first category, a prover and a verifier exchange multiple messages
(challenges and responses), typically dependent on random numbers
which they may keep secret whereas in the second the prover sends
only one message. In both systems the prover's objective is to
convince the verifier about the truth of an assertion, e.g. the claimed
knowledge of a secret. The verifier either accepts or rejects the
proof.

A zero-knowledge proof must obey the properties of completeness and
soundness. A proof is \textbf{complete}, if given an honest prover
and an honest verifier, the protocol succeeds with overwhelming
probability and \textbf{sound} if the probability of a dishonest
prover to complete the proof successfully is negligible
\cite{DBLP:conf/wecwis/AlmuhammadiSM04}. Additionally, a protocol
which consists a proof of knowledge must have the \textbf{zero-knowledge
property}: there exists an expected polynomial-time algorithm which
can produce, upon input of the assertions to be proven - but without
interacting with the real prover, transcripts indistinguishable from
those resulting from interaction with the real prover.

\begin{figure}
  \centering
    \subfigure[1st Step.]{
      \includegraphics[height=0.5in,width=0.7in]{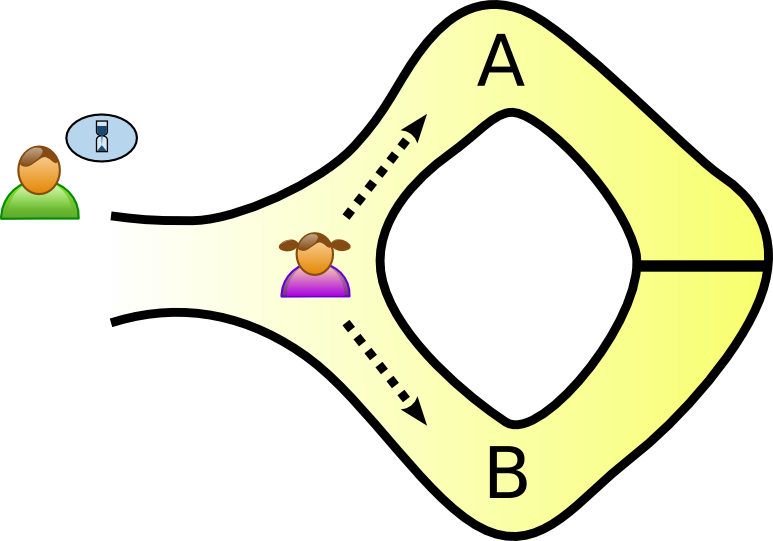}
      \label{fig:zkpalib1}
    }
    \hfill
    \subfigure[2nd Step.]{
      \includegraphics[height=0.5in,width=0.7in]{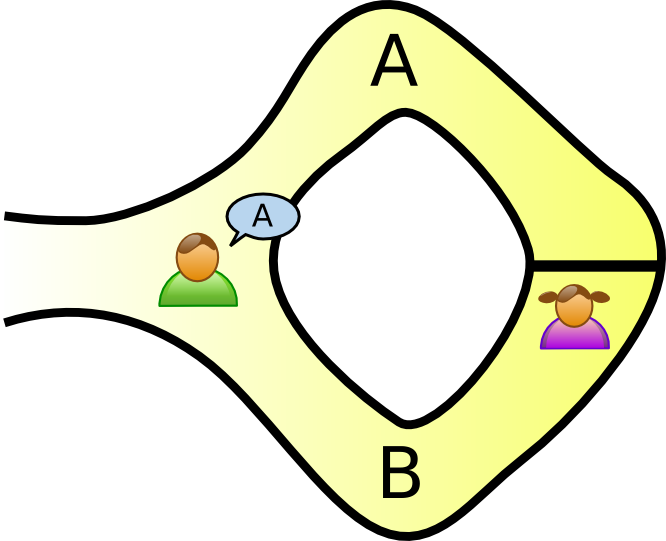}
      \label{fig:zkpalib2}
    }
    \hfill
    \subfigure[3rd Step.]{
      \includegraphics[height=0.5in,width=0.7in]{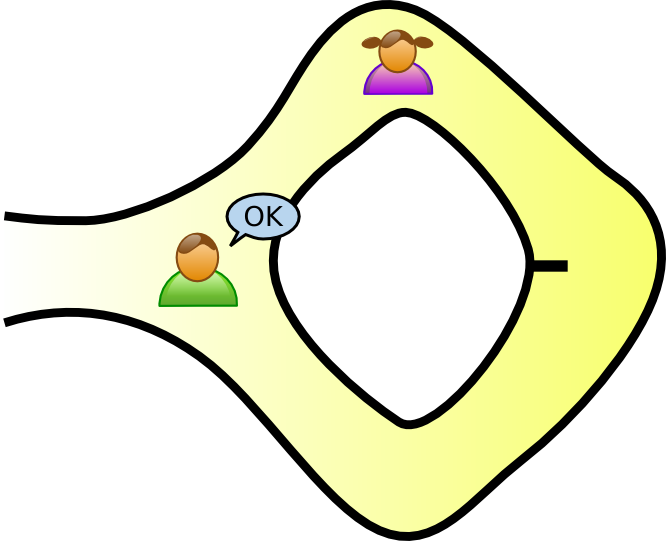}
      \label{fig:zkpalib3}
    }
  \caption{Alibaba's Cave Problem.}
  \label{fig:alib_cave}
\end{figure}

A typical example of zero-knowledge proof is known as Alibaba's cave
problem \cite{DBLP:conf/crypto/QuisquaterQQQGGGGGGB89}. In this
story, Peggy has uncovered the secret word used to open a \textit{magic} door
in a cave. The cave is shaped like a circle, with the entrance on
one side and the \textit{magic} door blocking the opposite side, as shown in
Figure~\ref{fig:alib_cave}. The left path from the entrance is
labeled A and the right B. Victor states that he will pay her for the
secret, but not until he's assured that she really knows it. Peggy
claims that she will tell him the secret, but not until she receives
the money. Thus, they devise a scheme by which Peggy can prove that
she knows the \textit{magic} word without telling it to Victor. The scheme steps
are now described:

\begin{itemize}
\item Victor waits outside the cave as Peggy goes in
\item Peggy randomly takes either path A or B inside the cave
\item Victor enters the cave and shouts the name of the path he wants her to use to return
either A or B, chosen at random
\item Peggy does that using the secret word if needed to open the magic door
\item The above steps are repeated n times until Victor is convinced that Peggy knows the
secret word
\end{itemize}

Now, suppose that Peggy does not know the secret word. Since Victor chooses
path A or B at random, Peggy has a $1/2$ chance of cheating at one
round. If the above steps are repeated for many rounds, Peggy's
chance of successfully anticipating all of Victor's requests would
become vanishingly small. Thus, if Peggy reliably appears at the
exit Victor names, he can conclude that she is very likely to know
the secret word.

Other problems that involve zero-knowledge proofs are the square
root of an integer modulo $n$, graph isomorphism, integer
factorization and the discrete logarithm problem. On this paper we
focus on zero-knowledge protocols based on the discrete logarithm
problem.
\section{Zero Knowledge Protocols based on the ECDLP}

A wide variety of zero-knowledge protocols based on the Discrete Logarithm Problem 
(DLP) has been proposed so far, e.g. in \cite{springerlink:10.1007/BF00196725}, \cite{Smith05cryptographymeets}.  
The Discrete Logarithm Problem is defined over arbitrary cyclic groups. A common example of cyclic group
is the multiplicative group $Z_{n}^{*}$ of order $n$, where $n$ is a prime number and the group operation
is multiplication modulo $n$. In such a group the Discrete Logarithm Problem (DLP) can be defined as follows:
Given a prime $n$, a generator $g$ of $Z_{n}^{*}$ and an element $b \in Z_{n}^{*}$, find the integer 
$x$, $ 0 \leq x \leq n-2$ such that $g^x = b (mod \, n)$ \cite{548089}. 

Another common example of cyclic groups are elliptic curve groups which are defined over
an additive group $F$ of order $n$ (note that $n$ is no longer necessarily a prime number).  
The analogous problem to DLP over elliptic curve groups is called ECDLP (Elliptic Curve Discrete Logarithm Problem) 
and can be defined as follows: Given an elliptic curve $E$ over a field $F$ of order 
$n$ (refered to as $F_{n}$ from now on), a generator point $G \in E / F_{n}$ and a point 
$B \in E/ F_{n}$ it is computationally hard to find $x$ such that $B = x \cdot G$. 

In this section, we show how well established zero-knowledge protocols based on the DLP 
can be adapted under the Elliptic Curve Discrete Logarithm Problem (ECDLP). This adaptation 
is a key step for porting such protocols to low constrained devices because of the Elliptic 
Curve Cryptography (ECC) advantages. As one can see in the \textbf{Appendix}, ECC can offer the 
same level of security as other public key cryptosystems, using smaller key sizes. This fact makes 
it suitable for implementations that concern constrained environments as it saves 
computational time and memory space and consequently reduces energy requirements. 
Such restrictions consist the real challenges when considering implementations on embedded devices.  

\subsection{Zero Knowledge Proof of Discrete Logarithm with Coin Flip}

One of the first zero-knowledge protocols of discrete logarithm that was originally 
presented in \cite{DBLP:conf/eurocrypt/ChaumEG87}. Its elliptic curve analogous
is as follows: Given an elliptic curve $E$ over a field $F_{n}$, a generator point 
$G \in E/ F_{n}$ and $B \in E/F_{n}$ Prover wants to prove that he knows $x$ such that 
$B = x \cdot G$, without revealing $x$.

\textbf{Protocol Steps:}
\begin{itemize}
\item Prover generates random $r \in F_{n}$ and computes the point $A = r \cdot G$
\item Prover sends the point $A$ to Verifier
\item Verifier flips a coin and informs the Prover about the outcome
\item In case of HEADS Prover sends $r$ to Verifier who checks that $r \cdot G = A$
\item In case of TAILS Prover sends $m = x + r (mod \, n) $ to Verifier who checks that $m \cdot G = (x+r) \cdot G = x \cdot G + r \cdot G = A + B$
\end{itemize}

The above steps are repeated until Verifier is convinced that Prover knows $x$ 
with probability $1 - 2^{-k}$ for $k$ iterations.

\textbf{Why it works:} The protocol works as expected because
in each iteration the steps to be executed depend on the outcome of
the coin that the Verifier flips and the Prover cannot affect this. 
It needs to be executed for many iterations in order for the Prover's cheating
probability to become very small. A dishonest Prover in each iteration can be 
prepared for only one of the coin outcomes and thus his cheating probability is $1/2$. 
For example, if he prepares for TAILS he can generate a random $m$,
compute $A = m \cdot G - B$ and send this point $A$ to Verifier. But
if HEADS come up this attack will not work. That is because he will need 
to compute a value $r \in F_{n}$ that generates $A$ and that is an instance
of the ECDLP. Thus, after $k$ iterations, the Verifier is convinced with high 
probability ($1 -2^{-k}$) that the Prover is honest. 

\subsection{Schnorr's Protocol}

An improvement of the previous protocol was originally presented in 
\cite{springerlink:10.1007/BF00196725}. The elliptic curve version
of Schnorr's protocol, slightly modified, is the following: 
Prover and Verifier agree on an elliptic curve $E$ over a field
$F_{n}$, a generator $G \in E/F_{n}$. They both know $B \in E/F_{n}$ 
and Prover claims he knows $x$ such that $B = x \cdot G$. 
He wants to prove this fact to Verifier without revealing $x$.

\textbf{Protocol Steps:}
\begin{itemize}
\item Prover generates random $r \in F_{n}$ and computes the point $A = r \cdot G$
\item Prover sends the point $A$ to Verifier
\item Verifier computes random $c = HASH(G, B, A)$ and sends $c$ to Prover
\item Prover computes $m = r + c \cdot x (mod \, n)$ and sends $m$ to Verifier
\item Verifier checks that $P = m \cdot G - c \cdot B = (r+c \cdot x) \cdot G -c \cdot B = r \cdot G + c \cdot x \cdot G - c \cdot x \cdot G = r \cdot G = A$
\end{itemize}

\textbf{Why it works:} This protocol is superior to the previous one as it needs to
be executed for one round. Verifier's coin flips (in correspondence with the Coin Flip protocol) 
are simulated using a hash function known only to him. A dishonest Prover has a tiny chance of 
cheating as he would have to fix the value of $P = m \cdot G - c \cdot B$ before receiving 
Verifier's hash value $c$. Under the assumption that the hash function used by the Verifier is 
secure, a Prover who does not know $x$, the discrete logarithm of $B$, cannot cheat.

\subsection{Transforming Schnorr's Protocol to Digital Signature}

In \cite{Fiat:1987:PYP:36664.36676}, the authors propose that with the use of a
hash function and an agreement on an initial message $m$ one can remove the 
interactivity from such protocols. The Verifier's random choices can be replaced 
with bits produced by a secure hash function. Thus, the next protocol is proposed.

Prover and Verifier agree on an elliptic curve $E$ over a field
$F_{n}$, a generator $G \in E/F_{n}$, a point $P \in E/F_{n}$ that
represents the message the Prover wants to send and a hash function
HASH (e.g. SHA-1). They both know $B \in E/F_{n}$. The Prover
claims that he knows $x$ such that $B= x \cdot G$ and he wishes to prove 
this fact to Verifier without revealing $x$.

\textbf{Protocol Steps:}
\begin{itemize}
\item Prover generates random $r \in F_{n}$ and computes the point $A = r \cdot G$
\item Prover computes $c = HASH(x \cdot P,r \cdot P, r \cdot G)$
\item Prover computes $s = r + c \cdot x (mod \, n)$
\item Prover sends to Verifier the message: ``$s || x \cdot P || r \cdot P || r \cdot G$''
\item Verifier computes $c= HASH(x \cdot P, r \cdot P, r \cdot G)$
\item Verifier checks that $s \cdot G = (r + c \cdot x) \cdot G = r \cdot G + c \cdot x \cdot G = r \cdot G + c \cdot B = A + c \cdot B$
\item Verifier checks that $s \cdot P = (r + c \cdot x) \cdot P = r \cdot P + c \cdot xP$
\end{itemize}

\textbf{Why it works:} In this protocol we apply the
non interactiveness trick proposed in \cite{Fiat:1987:PYP:36664.36676}. 
The Prover simulates both the Prover and the Verifier with the use of a hash 
function and publishes the transcript of this whole dialogue. This way the Prover
sends only one message and the Verifier either accepts or rejects. The
Prover generates a random number as in previous protocols but the
Verifier's random choices are simulated by hashing the input along with a
value calculated from the Prover's choice of $r$. Thus, the
Verifier's random choice depends on Prover's random choice and it is made hard to
fake the outcome. The value $c$ is really a challenge for the Prover
as it is computed from the hash function and it is out of his
control. If the Prover does not know $x$, in order to cheat he would
try to find $s$ satisfying $s \cdot G = r \cdot G + c \cdot x \cdot G$
which is an instance of the discrete logarithm problem. He could not cheat
by enumerating random $r$ values, as it would be too hard to find a
matching value for $c$.

\subsection{Zero Knowledge Test of Discrete Logarithm Equality}

Suppose that Prover knows two publically known quantities that have
the same discrete logarithm $x$ to publicly known respective bases
$G$ and $H$ of the group $F_{n}$.

Prover and Verifier agree on an elliptic curve $E$ over a field $F_{n}$, 
a generator $G \in E / F_{n}$ and $H \in E/F_{n}$. Prover claims he knows 
$x$ such that $B = x \cdot G$ and $C = x \cdot H$ and wants to prove knowledge 
of this fact without revaling $x$. The procedure
was originally proposed in \cite{DBLP:conf/crypto/BoyarCDP90}, and its 
ECC analogous is as follows:

\textbf{Protocol Steps:}
\begin{itemize}
\item Prover chooses random $r \in F_{n}$ and computes the points $K = r \cdot G$ and $L = r \cdot H$
\item Prover sends the points $K, L$ to Verifier
\item Verifier chooses random $c \in F_{n}$ and sends $c$ to Prover
\item Prover computes $m = r + c \cdot x (mod \, n)$ and sends $m$ to Verifier
\item Verifier checks that $m \cdot G = (r + c \cdot x) \cdot G = r \cdot G + c \cdot x \cdot G = K + c \cdot B$
\item Verifier checks that $m \cdot H = (r + c \cdot x) \cdot H = r \cdot H + c \cdot x \cdot H = L + c \cdot C$
\end{itemize}

\textbf{Why it works:} In this protocol the Prover claims he knows
$x$ as the discrete logarithm of two public quantities $B, C$. His actions
are similar with Schnorr's protocol but for the two public
quantities. For example in the first step he computes 2 points on
the curve $K, L$ that will be used for the verification. It can also
be made non-interactive by the applying the Fiat-Shamir trick: 
the Prover simulates the Verifier by computing $c$ with a secure hash 
function as HASH(B,G,C,H,K,L).

\subsection{Zero Knowledge Proof of Single Bit}

Prover and Verifier agree on an elliptic curve $E$ over a field
$F_{n}$, a generator $G \in E / F_{n}$ and $H \in E/F_{n}$. Prover
knows $x$ and $h$ such that $B = x \cdot G + h \cdot H$ where $h =
\pm 1$. He wishes to convince Verifier that he really does
know $x$ and that $h$ really is $\pm 1$ without revealing $x$ nor the sign bit 
\cite{Smith05cryptographymeets}.

\textbf{Protocol Steps:}
\begin{itemize}
\item Prover generates random $s,d,w \in F_{n}$
\item Prover computes the points $A = s \cdot G - d \cdot (B+ h \cdot H)$ and $C = w \cdot G$
\item If $h = -1$ Prover swaps $A \leftrightarrow C$
\item Prover sends the points $A,C$ to Verifier
\item Verifier generates random $c \in F_{n}$ and sends $c$ to Prover
\item Prover computes $e = c - d$ and $ t = w + x \cdot e$ both $(mod \, n)$
\item If $h = -1$ Prover swaps $d \leftrightarrow e$ and $ s \leftrightarrow t$
\item Prover sends to Verifier $d, e, s, t$
\item Verifier checks that $e + d = c$, $s \cdot G = A + d \cdot (B+H)$ and that $t \cdot G = C + e \cdot (B-H)$
\end{itemize}

\textbf{Why it works:} It is straightforward to confirm that if $B$
is really given by one of the two formulas the Prover claimed then
Verifier's verification will succeed. It is also easy to see that
the Prover does not give away any information that would allow the
Verifier to deduce $x$ nor the sign bit $h$. That is because $x$ is
hidden inside $t$ after being multiplied with $e$ and added in $w$.
The sign bit $h$ is randomized with the appropriate swaps in the
case of $-1$.
\section{Protocols Implementation and Evaluation}

We implemented the new proposed zero-knowledge protocols based on the ECDLP using
the Wiselib platform which is a generic algorithm library. Then, we evaluated the 
implemented protocols on two popular hardware platforms equipped with 
popular low-end microcontrollers (Jennic JN5139, TI MSP430) as well as 802.15.4 RF transceivers
in terms of execution time, code size, message size and energy consumption.

\subsection{Wiselib: A Generic Algorithm Library for Sensor Networks}

We decided to implement our algorithms using \textbf{Wiselib}
\cite{DBLP:conf/ewsn/BaumgartnerCFKKP10}: a code library, that 
allows implementations to be OS-independent. It is implemented based on 
C++ and templates, but without virtual inheritance and exceptions. Algorithm 
implementations can be recompiled for several platforms and firmwares, without 
the need to change the code. \textbf{Wiselib} can interface with systems implemented 
using C (Contiki), C++ (iSense), and nesC (TinyOS). A future plan for this library
is to be adapted for C-based mobile phone operating systems like
Android and iPhone OS.

Furthermore, an important feature of \textbf{Wiselib} are the already
implemented algorithms and data structures. Since different kind of
hardware uses different ways to store data (due to memory alignment,
inability to support dynamic memory, etc.), it is important to use these
safe types as much as possible since they have been tested before on most
hardware platforms. As of mid 2010, the \textbf{Wiselib} includes about 40
Open Source implementations of standard algorithms, and is scheduled to
grow to 150-200 algorithms by the end of 2011.

Additionally, \textbf{Wiselib} runs on the simulators
\textbf{Shawn}~\cite{DBLP:journals/corr/abs-cs-0502003} and \textbf{Tossim}~\cite{DBLP:conf/sensys/LevisLWC03}, 
hereby easing the transition from simulation to actual devices. 
\textbf{Tossim} is a popular tool in the TinyOS
community as it allows to simulate the exact source code that will run on the hardware
and by using \textbf{PowerTossim}~\cite{Shnayder:2004:SPC:1031495.1031518} it can provide
accurate estimates on the power consumption of an application. 
\textbf{Shawn} allows repeatability of simulations in an easy way by using only a single
configuration file. It provides many options such as packet loss, radius
of communication, ways of communicating and even mobility in
an abstract way, without needing to provide specific code for every range.
This \textbf{Wiselib} feature allows us to validate the faithfulness of our implementation 
and also get results concerning the quality of our algorithms without time 
consuming deployment procedures and harsh debugging environments.

Finally, an advantage of Wiselib is that with the aid of template
specializations the algorithm code can be optimized and adapted for
certain platforms. Depending on the compilation process, the
compiler can select the code that fits best for the current platform
(e.g. if there is a 32-bit processor) and exploit the presence of
special platform hardware (e.g. the Jennic AES hardware for speedup
of crypto routines).

\subsection{Hardware}

As mentioned earlier we used two different low-end devices for evaluating
the implemented protocols. The first device is the Coalesenses iSense
\cite{buschmann07isense}, \cite{coalesenses} and the second is the Crossbow
TelosB \cite{crossbow}. These devices are quite popular for
their application in the area of wireless sensor networks.

The first device (iSense) consists of a Jennic JN5139 32-bit RISC controller
\cite{jennic} running at 16MHz. The ROM of this controller is 192Kb and its RAM is 96Kb
that can be shared among program and data. It is equipped with 2.4Ghz IEEE 802.15.4 compliant 
RF transceiver (CC2420 chip) that can achieve bandwith up to 250 Kbps. Finally, this device runs 
the iSense firmware.

The second device (TelosB) consists of a Texas Instruments MSP430 \cite{Davies:2008:MMB:1481318} 16-bit
microcontroller running at 8MHz. Its RAM is 10Kb and the program
flash memory is 48Kb. This device is also equipped with 2.4GHz IEEE 802.15.4 compliant
RF transceiver (CC2420 chip) able to achieve data rates up to 250Kbps. Finally, the TelosB device 
can run TinyOs 1.1.10 \cite{tinyos} (or higher) or the Contiky operating system \cite{contiki}.

\begin{figure}
  \centering
    \subfigure[A TelosB device.]{
      \includegraphics[height=0.8in, width=1.1in]{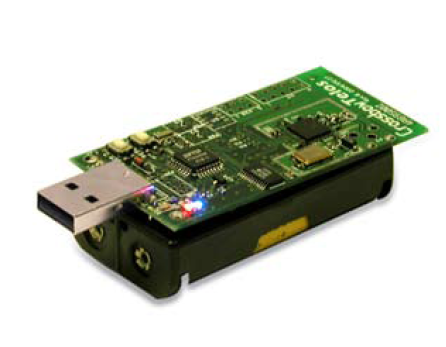}
      \label{fig:telosb_}
    }
    \hfill
    \subfigure[An iSense device.]{
      \includegraphics[height=0.6in, width=0.9in]{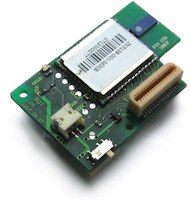}
      \label{fig:isense_}
    }
  \caption{The hardware used for our experiments.}
  \label{fig:hardware}
\end{figure}

\subsection{Results}

For the basic operations of Elliptic Curve Cryptography we have
ported the implementation of \cite{DBLP:journals/tosn/MalanWS08} on
the Wiselib. This implementation defines a recommended elliptic
curve \cite{sec2} over binary fields with equation $y^2 + xy = x^3 + x^2 + 1$
along with the irreducible polynomial $f(x) = x^{163} + x^7 + x^6 +
x^3 + 1$. The curve's order (the number of points on it) is $r =
0x4000000000000000000020108a2e0cc0d99f8a5ef$ and the base point is
$G (x,y)$ where \\ $x = 0x2fe13c0537bbc11acaa07d793de4e6d5e5c94eee8$
and \\ $y = 0x289070fb05d38ff58321f2e800536d538ccdaa3d9$. The
execution time of the basic elliptic curve operations for both the
microprocessors used can be seen on Table \ref{exec_time1}. We note
that there has been no attempt to optimize the code responsible for
the elliptic curve operations.

\begin{center}
\begin{table}[h]
\begin{scriptsize}
  \begin{tabular}{ |  m{4.2cm} ||  c | c | }
    \hline
    Operation & \multicolumn{2}{ | c |}{Execution Time}\\
    & JN5139 & MSP430 \\ \hline
    Private Key Generation &  0.087 sec & 0.3 sec\\ \hline
    Public Key Generation (scalar multiplication) & 11.121 sec & 58.02 sec \\ \hline
    Curve Points Addition &  0.094 sec & 0.29 sec\\ \hline
  \end{tabular}
\caption{Execution Time of Basic Elliptic Curve Operations on the JN5139 and MSP430 microprocessors.}
\label{exec_time1}
\end{scriptsize}
\end{table}
\end{center}

Table \ref{exec_time2} summarizes the execution times of some additional arithmetic
operations used by the protocols on the JN5139 and MSP430 microcontrollers.
As a hash function for the protocols that require one we used the
algorithm SHA-1 \cite{citeulike:1576034}.

\begin{center}
\begin{table}[h]
\begin{scriptsize}
  \begin{tabular}{ | c || c | c | }
    \hline
    Operation & \multicolumn{2}{| c |}{Execution Time} \\
    & JN5139 & MSP430 \\ \hline
    Private Key Addition (21 bytes) &  0.005 sec & 0.012 sec\\ \hline
    Private Key Multiplication (21 bytes) & 0.014  sec & 0.02 sec \\ \hline
    SHA-1 Hash (250 bytes) &   0.02 sec & 0.031 sec \\ \hline
\end{tabular}
\caption{Execution Time of Arithmetic Operations Used by the Protocols on the JN5139 and MSP430 microprocessors.}
\label{exec_time2}
\end{scriptsize}
\end{table}
\end{center}

From these experiments, a reader can observe that the elliptic curve scalar multiplication
(Public Key Generation) is the most demanding arithmetic operation.
Moreover, it is also evident that the JN5139 (16MHz) calculates much
faster all the operations examined than the MSP430 (8MHz). We believe that a delay of 11 sec for the generation
of a public key is acceptable. On the other hand, the 58 sec that are required from
the MSP430 is probably too long.

For all the arithmetic operations previously mentioned, we present a
theoretical approach for estimating their energy
consumption. This approach is based on the current consumption that
the data sheet of each microprocessor provides us. For example, from the
graphs presented in JN5139 data sheet we get that
the current consumption of the mote when the microprocessor works on
maximum load is 12.7mA, on input voltage 3V and temperature
$25^{\circ}$C. The input voltage 3V is a reasonable choice
considering the fact that two AA batteries provide that much
voltage. Respectively, the MSP430 microcontroller draws
approximately 1.8mA current when it operates. Thus, considering the
formula
\begin{equation}
\label{energ1}
E = V \cdot I \cdot t
\end{equation}
where $V$ is voltage, $I$ is current and $t$ is time, and the execution time of each operation, as seen on
Tables \ref{exec_time1} and \ref{exec_time2},
we show on Table \ref{energ} our estimation for each operation's energy consumption.

\begin{center}
\begin{table}[h]
\begin{scriptsize}
  \begin{tabular}{ | m{4.5cm} ||  c | c | }
    \hline
    Operation & \multicolumn{2}{ | c |}{Energy Consumption}\\
    & JN5139 & MSP430 \\ \hline
    Private Key Generation &  3.31 mJ & 1.62 mJ \\ \hline
    Public Key Generation (scalar multiplication) &  423.6 mJ & 313.3 mJ \\ \hline
    Curve Points Addition &  3.42 mJ & 1.56 mJ \\ \hline
    Private Key Addition (21 bytes) & 0.19 mJ  & 0.06 mJ \\ \hline
    Private Key Multiplication (21 bytes) & 0.53 mJ & 0.1 mJ \\ \hline
    SHA-1 Hash (250 bytes) &  0.76 mJ & 0.16 mJ \\ \hline
  \end{tabular}
\caption{Energy Consumption of Basic Arithmetic Operations Used by the Protocols on the JN5139 and MSP430 microprocessors.}
\label{energ}
\end{scriptsize}
\end{table}
\end{center}

As expected, on Table \ref{energ}, we observe that the elliptic curve scalar multiplication is the most
energy consuming arithmetic operation. Additionally, it is interesting to note that the JN5139 processor, 
although it is much faster in computation than the MSP430, it is less efficient in terms of energy consumption.  

Next, we have concentrated the prover's (PRV) and verifier's (VER) actions 
for each of the implemented protocols. Table \ref{actions} summons up
these actions. Regarding to these actions we verify the total protocol
execution times that can be seen subsequently.

\begin{table}[h]
\begin{tiny}
  \begin{tabular}{ | m{1cm} || m{.5cm} | m{.45cm} | m{.45cm} | m{.43cm} | m{.5cm} | m{.5cm} | m{.42cm} |}
    \hline
    Protocol / Operation & Random Key Generation & Curve Multiplication & Curve Addition & SHA-1 Hash & Private Keys Addition & Private Keys Multiplication & Msgs Sent\\ \hline
    ZKP of DL with Coin Flip PRV & 1 & 1 & - & - & 1 (if tails) & - & 2 \\ \hline
    ZKP of DL with Coin Flip VER & - & 1 & 1 (if tails) & - & - & - & 2 \\ \hline
    Schnorr's Protocol PRV & 1 & 1 & - & - & 1 & 1 &  2 \\ \hline
    Schnorr's Protocol VER & - & 2 & 1 & 2 & - & - & 2 \\ \hline
    Schnorr's Signature Protocol PRV & 1 & 3 & -& 1 & 1 & 1 & 1 \\ \hline
    Schnorr's Signature Protocol VER & - & 4 & 2 & 1 &- & -& 1 \\ \hline
    ZKP of DL Equality PRV & 1 & 2 & -& -& 1& 1& 2 \\ \hline
    ZKP of DL Equality VER & 1 & 4 & 2 & - & - & - & 2 \\ \hline
    ZKP of Single Bit PRV & 3 & 3 & 2 & - & 2 & 1 & 2 \\ \hline
    ZKP of Single Bit VER & 1 & 4 & 4 & - & 1 & - &2  \\ \hline
  \end{tabular}
\caption{Actions Held by the Prover (PRV) and Verifier (VER) for each protocol.}
\label{actions}
\end{tiny}
\end{table}

On Table \ref{totalexectime} one can see the total execution time of the
above implemented protocols. This is measured as the time space
between the prover's beginning of the protocol until the verifier's
final response of whether he accepts or rejects the proof. We can observe that
an interactive protocol like the first one, which requires a large
number of execution rounds for verification, is not suitable for
low-constrained devices. Thus, the usage of
protocols (like the rest of them) that need one round of execution
is advised \cite{DBLP:conf/wecwis/AlmuhammadiN05}.

\begin{center}
\begin{table}[h]
\begin{scriptsize}
  \begin{tabular}{ | m{2.1cm} || c |  c | c | }
    \hline
    Protocol & Required Rounds & \multicolumn{2}{|c|}{Total Execution Time} \\ \hline
    & & iSense & TelosB \\
    ZKP of DL with Coin Flip & 100 or more &  2277 sec & 11802 sec\\ \hline
    Schnorr's Protocol & 1 & 33.894 sec & 172.77 sec \\ \hline
    Schnorr's Signature Protocol & 1 & 78.645 sec & 396.57 sec \\ \hline
    ZKP of DL Equality & 1 & 68.596 sec & 346.2 sec \\ \hline
    ZKP of Single Bit & 1 & 80.46 sec & 462.74 sec \\ \hline
  \end{tabular}
\caption{Total Execution Time of the Protocols on the Devices Used.}
\label{totalexectime}
\end{scriptsize}
\end{table}
\end{center}

Table \ref{messagesize} describes the messages and their sizes, exchanged by the prover (PRV)
and verifier (VER) for the completion of each protocol. Message size is an important parameter
when considering low power wireless communication protocols like 802.15.4. All of our protocols 
messages except Schnorr's Non-Interactive Protocol PRV message (149 bytes which had to be broken in
two pieces), fit well on a single 802.15.4 packet (128 bytes). This fact is advantageous as bigger 
messages would result to more message exchanges which in turn would
result to greater energy consumption and possibly to wireless medium congestion. 

\begin{center}
\begin{table}[h]
\begin{scriptsize}
  \begin{tabular}{ | m{3.5cm} ||  m{3.5cm} | }
    \hline
    Protocol & Message Size \\ \hline
    \multirow{2}{*}{ZKP of DL with Coin Flip PRV} &  Point Message: 43 bytes \\
     & New Key Message: 22 bytes \\ \hline
    \multirow{2}{*}{ZKP of DL with Coin Flip VER}& Coin Message: 2 bytes \\
     & Final Message: 1 byte \\ \hline
    \multirow{2}{*}{Schnorr's Protocol PRV} & Point Message: 43 bytes \\
      & New Key Message: 22 bytes \\ \hline
    \multirow{2}{*}{Schnorr's Protocol VER} & Hash Message: 22 bytes \\
      & Final Message: 1 byte \\ \hline
    Schnorr's Signature Protocol PRV & Point and Key Message: 149 bytes (in two pieces) \\ \hline
    Schnorr's Signature Protocol VER & Final Message: 1 byte \\ \hline
    \multirow{2}{*}{ZKP of DL Equality PRV} & Points Message: 85 bytes \\
      & New Key Message: 22 bytes \\ \hline
    \multirow{2}{*}{ZKP of DL Equality VER} & New Key Message: 22 bytes \\
    & Final Message: 1 byte \\ \hline
    \multirow{2}{*}{ZKP of Single Bit PRV} & Points Message: 85 bytes \\
      & New Key Message: 85 bytes \\ \hline
    \multirow{2}{*}{ZKP of Single Bit VER} & New Key Message: 22 bytes \\
    & Final Message: 1 byte \\ \hline
\end{tabular}
\caption{Size of Messages Exchanged by the Prover (PRV) and the Verifier (VER) for each Protocol.}
\label{messagesize}
\end{scriptsize}
\end{table}
\end{center}

It is quite difficult to estimate the energy consumption of RF
messages as they depend on various factors like motes distance,
obstacles presence and weather conditions. However, we try to make a
theoritical estimation about it. Under the \textit{ideal} hypothesis
that the devices' RF transceivers achieve data rate 250 Kbps and
according to Table \ref{messagesize} we calculate the time required to
send each message depending on its size (e.g. with data rate 250 Kbps a 
message of size 85 bytes needs around
2.72 msec for transmission/reception). Next, we estimate the
energy consumption for sending and receiving the messages on each
device using the formula \ref{energ1}. From the iSense data sheet we get that the Rx current
consumption on input voltage 3V and temperature $25^{\circ}$C is
43.7 mA and the Tx current consumption is 39.9 mA. Respectively,
from the TelosB data sheet we get that Rx current consumption is
23mA and that Tx current consumption is 21mA. Table \ref{messages}
summarizes the results.

\begin{center}
\begin{table}[h]
\begin{scriptsize}
  \begin{tabular}{ | c || c | c| c |  c | }
    \hline
    Message Size & \multicolumn{2}{|c|}{iSense} & \multicolumn{2}{|c|}{TelosB} \\
     & Tx & Rx & Tx & Rx \\ \hline
    149 bytes & 0.5 mJ & 0.6 mJ & 0.3 mJ & 0.32 mJ \\ \hline
    85 bytes & 0.32 mJ & 0.36 mJ & 0.17 mJ & 0.18 mJ \\ \hline
    43 bytes & 0.16 mJ & 0.18 mJ & 0.08 mJ & 0.09 mJ \\ \hline
    22 bytes & 0.08 mJ & 0.09 mJ & 0.045 mJ & 0.048 mJ \\ \hline
    2 bytes & 0.007 mJ & 0.008 mJ & 0.004 mJ & 0.004 mJ \\ \hline
    1 byte & 0.003 mJ & 0.004 mJ & 0.002 mJ & 0.002 mJ \\ \hline
  \end{tabular}
\caption{Message Tx, Rx Energy Consumption According to its Size on the Devices Used.}
\label{messages}
\end{scriptsize}
\end{table}
\end{center}

In correspondence with the Tables
\ref{energ}, \ref{actions}, \ref{messages} we estimate the total
energy consumption for the prover and verifier of each protocol
depending on its actions (arithmetic operations and messages
sent/received). Table \ref{totalenerg} shows the results. Once again, we realize
why a protocol (like ZKP of DL with Coin Flip) that requires many
execution rounds, is not suitable for applications that involve
constrained devices. Moreover, we observe that the protocols execution consumes less
energy (although more time) on the TelosB device with the slower processor. 

\begin{center}
\begin{table}[h]
\begin{scriptsize}
  \begin{tabular}{ | m{3cm} || m{0.8cm} | m{0.8cm} | }
    \hline
    Protocol & \multicolumn{2}{|m{2.7cm}|}{Total Energy Consumption} \\
     & iSense & TelosB \\ \hline
    ZKP of DL with Coin Flip PRV & 42.7 J & 31.5 J \\ \hline
    ZKP of DL with Coin Flip VER & 42.5 J & 31.4 J \\ \hline
    Schnorr's Protocol PRV & 0.42 J & 0.31 J \\ \hline
    Schnorr's Protocol VER & 0.85 J & 0.62 J\\ \hline
    Schnorr's Signature Protocol PRV & 1.27 J & 0.94 J \\ \hline
    Schnorr's Signature Protocol VER & 1.7 J & 1.25 J \\ \hline
    ZKP of DL Equality PRV & 0.85 J & 0.62 J \\ \hline
    ZKP of DL Equality VER & 1.7 J & 1.25 J\\ \hline
    ZKP of Single Bit PRV &  1.28 J & 0.94 J \\ \hline
    ZKP of Single Bit VER & 1.71 J & 1.26 J \\  \hline
  \end{tabular}
\caption{Total Energy Consumption for the Prover (PRV) and the Verifier (VER) of each Protocol on the Devices Used.}
\label{totalenerg}
\end{scriptsize}
\end{table}
\end{center}

From tables \ref{totalexectime} and \ref{totalenerg}, we note that Schnorr's protocol is the fastest one and thus consumes the
least energy. We think that 33 sec for the completion of a zero-knowledge proof on a device with limited processing power is quite fair. 

As a last experiment, we measured the code size of the above protocols on the devices used. 
The compiled code actually fits fairly well (approximately 8Kb) on the tiny memory of their processors. 
The results can be seen on Table \ref{codesize}. The difference in the protocols code size on the two devices
is reasonable as different compilers (ba-elf for iSense and mspgcc for TelosB) are employed.

\begin{center}
\begin{table}[h]
\begin{footnotesize}
  \begin{tabular}{ | m{3.2cm} || m{1.5cm} | m{1.5cm} | }
    \hline
    Protocol &\multicolumn{2}{|m{3.5cm}|}{Total Code Size (text + data + bss)}  \\ \hline
     & iSense & TelosB \\
    ZKP of DL with Coin Flip PRV &  7908 bytes & 6517 bytes \\ \hline
    ZKP of DL with Coin Flip VER &  7004 bytes & 6289 bytes\\ \hline
    Schnorr's Protocol PRV & 7964 bytes & 6759 bytes\\ \hline
    Schnorr's Protocol VER & 9292 bytes & 9455 bytes\\ \hline
    Schnorr's Signature Protocol PRV & 10584 bytes & 10433 bytes\\ \hline
    Schnorr's Signature Protocol VER & 9540 bytes & 10053 bytes\\ \hline
    ZKP of DL Equality PRV & 8568 bytes & 7697 bytes \\  \hline
    ZKP of DL Equality VER & 8964 bytes & 8519 bytes \\  \hline
    ZKP of Single Bit PRV & 9604 bytes & 9471 bytes \\  \hline
    ZKP of Single Bit VER & 9032 bytes & 7455 bytes \\  \hline
  \end{tabular}
\caption{Code Size of the Protocols for Prover the (PRV) and the Verifier (VER) on the Devices Used.}
\label{codesize}
\end{footnotesize}
\end{table}
\end{center}

\subsection{Discussion}

From the results presented in the previous subsection we stick to the most important observations.  

First of all, we state why implementing the ZKP protocols under the elliptic curve cryptography setting is 
advantageous. As presented in the \textbf{Appendix}, ECC offers the same level of security as other public 
key cryptosystems, using smaller key sizes. An implementation of zero knowledge protocols of discrete logarithm
over multiplicative groups would require the use of at least 1024-bit keys as proposed by NIST (see Figure~\ref{fig:key_equiv} in Appendix).
A reader can realize that using elliptic curve groups instead of multiplicative groups, the arithmetic operations need less
time to execute, less memory space and thus less energy consumption. In fact, it has been proven that ECC actually outperforms RSA 
on constrained environments in terms of computation time, memory requirements and thus energy consumption \cite{citeulike:2336173}. 
Moreover, considering the wireless nature of communication on the embedded setting ECC offers the advantage of smaller message
sizes that cost less and have better chances of being delivered. 

Concerning the protocols performance we discuss the experimental results.

The first proposed protocol (ZKP of DL with Coin Flip) requires a large number of execution rounds in order to 
successfully complete. It performs a large number of arithmetic operations and it involves 
many message exchanges. Although such a protocol could be executed fairly in non-embedded systems, when
considering embedded devices capable of wireless communication it is not suitable. It is very demanding in
terms of energy consumption and the messages exchanged can congest the wireless medium.
The rest of the protocols require only one round of execution \cite{DBLP:conf/wecwis/AlmuhammadiN05}, much smaller number of messages 
(2-4 messages) and complete much faster and with less energy consumption. Schnorr's protocol required the least time to execute - 33 sec 
on the JN5139 microcontroller, which is quite fair considering the processing power and the memory of the device used. 

An additional disadvantage of the first protocol is that it cannot be considered as secure as the rest
when dealing with wireless communication. A malicious verifier or an adversary who eavesdrops the communication between 
the prover and the verifier can replay the proof to another party using the overheard data. That is because 
the verifier's responses consist of just a single bit (simulating a coin flip). In all the other protocols 
the Verifier responses involve fresh random data (e.g. using a hash function) and such an attack could not work. 

Another issue concerns the comparison between Schnorr's and Schnorr's non-interactive protocol. By transforming 
Schnorr's protocol to a non-interactive protocol, the required message exchanges are reduced (1 message required by the 
prover and verifier). However, the transformed protocol is not as efficient as Schnorr's original protocol in terms 
of execution time, energy consumption, code size and message size (see below).  

As far as the efficiency of the elliptic curve arithmetic operations of our implementation is concerned,
we have already mentioned that we did not make any attempts to optimize the code. 
One can easily observe that the curve point multiplication which is the basic action for each
protocol, needs the most time to execute (11 sec on JN5139, 58 sec on MSP430). Subsequently, this 
action requires the most energy for its execution. We believe that this is the \textit{price} you pay
when you write some generic code which can be portable. Although, Wiselib offers the chance to compile your
code exploiting some platform features (a 32-bit processor or hardware speedups) we aimed at generality and 
portability. The goal of this work was not to achieve some faster platform specific code
(e.g. by employing low-level assembly code) as in \cite{citeulike:2336173}, \cite{DBLP:conf/ipsn/LiuN08}.

However, we think that for such low-end microprocessors running at
16 MHz or 8MHz the total execution time of our protocols (except of course ZKP of DL with Coin Flip) is
reasonable. For example, 33 sec on a 16MHz microcontroller for a secure verification is not too much
to wait. Having implemented our protocols in Wiselib, with little effort we can port our code on C-based 
operating systems for mobile phones like Android and iPhone OS where the embedded hardware is much more powerful. 
For instance, we note that a current HTC mobile phone is powered by an 1 GHz ARMv7 Snapdragon processor
with 512 MB of flash and 576 MB of RAM. Of course, our code would execute much faster on 
such a platform. Nevertheless, running and evaluating it on really low-end devices was much more meaningful.

Considering memory limitations, we showed that the protocols' compiled code actually fits well on the
restricted memory of the devices used. For the iSense devices the protocols code occupied approximately 
9Kb out of the total 96Kb of memory that is offered. Respectively, the protocols code on the TelosB devices, 
occupied approximately 8Kb out of the total flash memory of 48Kb.

Moreover, as far as the protocols' message sizes are concerned, we observe that they are acceptable. 
Most of the messages exchanged are 43 or 85 bytes which can fit on a single 802.15.4 packet (max payload 128 bytes). 
Only, the Prover's message from Schnorr's non-interactive protocol was large enough (149 bytes)
that could not fit in a single packet and thus had to be broken in two pieces. However, by recompiling 
each device's firmware, which is a straightforward procedure, one could increase the maximum 
payload size and such a message could fit in a single packet. Such a solution though, would be non-standard. 
The reasonable message sizes are due to the ECC approach. A reader can realize that using 
another cryptosystem (e.g. with 1024-bit keys) would result in larger message sizes that would require
fragmentation. This way many messages would need to be exchanged and more energy would be consumed. 
Additionally, the possibility of communication faults would increase as the wireless medium could be congested.

Finally, a point for discussion, is the fact that in our protocols we preload the elliptic curve used and 
its parameters on the devices memory. This way, the elliptic curve and its parameters could be compromised by 
physical and tampering attacks. However, we believe that in the future, most devices will be
equipped with Trusted Platform Module (TPM), secure cryptoprocessors suitable for storing cryptographic keys 
and protecting private information.
\section{Application Scenarios}

We are strongly confident that zero-knowledge proofs can be used as a privacy-preserving tool
in future global networks consisting of different kinds of devices and thus we present some 
everyday's application examples:

\subsection{Course Polling at University}

In most universities of the industrial world, course evaluations have become standard practise but
they are typically conducted outside computers in order to protect student's privacy. An application
that would encourage students to anonymously state their opinion about a course is descirbed.

Imagine a concept where students can provide feedback about a course or a lecture using tiny devices capable 
of communicating wirelessly (e.g. clickers). For the success of this concept two parameters should be considered. 
Firstly, only authorized students (students who actually attend the course) should be able to poll and
secondly students' anonymity should be guaranteed. Thus, when a student enrols for a course at the University's
Administration Office (a trusted authority) he gets a tiny device that holds a private key encoding some of his
attributes (e.g. name, matriculation number, course id and some fresh random value). This private key is used to generate 
a public key which is stored by the authority on a list. This list is then provided to the professor who is responsible for the course.
At the end of a lecture the professor activates a polling application on his laptop (equipped with RF transceiver) 
that presents questions to the students. The students can now answer the poll questions using their devices. The application 
accepts a student's answer if the student's device presents a public key that exists on the list of valid keys (as provided by the 
University's Administration Office) and proves that it is the actual owner of the private key that has generated it. This way, 
student's anonymity is preserved (his private key is never revealed) and the professor gets the poll results while being assured 
that students who do not attend the class or malicious entities who eavesdrop the messages exchanged during the polling 
procedure are not able to poll.

For the deployment of such an application, a zero-knowledge protocol of discrete logarithm is required. Regarding the 
performance evaluation results that were presented in the previous section we propose \textit{Schnorr's Protocol}. 

\subsection{Anonymous Travel Ticket plus Discount Benefits}

\begin{figure}
  \centering
    \subfigure[1st Step.]{
      \includegraphics[height=0.9in,width=1.37in]{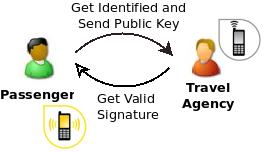}
      \label{fig:port1}
    }
    \hfill
    \subfigure[2nd Step.]{
      \includegraphics[height=0.9in,width=1.37in]{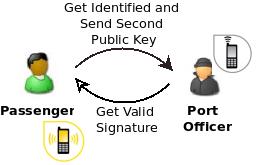}
      \label{fig:port2}
    }
    \hfill
    \subfigure[3rd Step.]{
      \includegraphics[height=0.9in,width=1.37in]{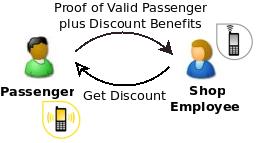}
      \label{fig:port3}
    }
  \caption{Boat Travel Ticket plus Discount Benefits Application.}
  \label{fig:boat_app}
\end{figure}

Nowadays, it is quite common for passenger ships that travel over national waters to have duty free shops in them (e.g. boats
that travel on the Baltic Sea). When passengers enter these shops, they are usually requested to present their ticket in order to buy
some goods. As travel tickets typically contain a passenger's name, privacy issues are raised. For example, a passenger may not
desire for his purchases to be publically known e.g. for statistical purposes. However, a passenger that has the benefit of some 
discount (e.g. a first class passenger) needs a way to prove this fact without revealing his private data. We present an application
that can deal with these issues (Figure~\ref{fig:boat_app}).

\textbf{First Step:} A passenger visits a travel agency office in order to buy a boat ticket. As he purchases the ticket, his mobile phone which is 
able to communicate wirelessly through Near Field Communications (NFC) stores a private key that encodes his attributes 
(name, travel class and some random fresh value). This private key is used to generate a public key that is then digitally signed by 
the travel agency with a secret key depending on the passenger's travel class. 

\textbf{Second Step:} When passengers approach the port entrance, they identify themeselves to the port officer as proposed by international travel safety regulations. 
Once a passenger that travels first class (and desires the right to some discount benefits on the ship shops) is identified 
by the port officer, he generates a second public key that encodes his right to discount with the same private key using his mobile phone. 
The second key is digitally signed by the port officer's device. 

\textbf{Third Step:} From now on, a passenger is not required to identify himself at any occasion on the boat. When a passenger enters the ship his mobile phone 
communicates with the device of the ticket control employee and proves that he is eligible to travel at the corresponding class without revealing 
his private data (using a zero-knowledge proof of discrete logarithm e.g. \textit{Schnorr's protocol}). When a first class passenger who has 
additionally discount benefits wants to purchase some products he enters the duty free shop and his mobile phone communicates through NFC with 
the employee's device and proves two things: first that he is a valid first class passenger and second that he has the right to product discount. 
This proof takes place by showing that two digitally signed by the corresponding authorities public keys have the same discrete logarithm 
(use of the protocol \textit{Zero Knowledge Test of Discrete Logarithm Equality}). If the proof succeeds the passenger gets a discount on his purchases. 
A malicious user, is not able to extract information about passenger's private data (e.g. his name) or to get product discount if he does not 
match the necessary criteria.

\subsection{Parking in Smart Cities}

Finally, we show how the protocol \textit{Zero-Knowledge Proof of Single Bit} can be used in an application. This protocol proves the possesion
of a discrete logarithm without revealing it and additionally proves that a value has been added or substracted to the 
corresponding public key without revealing which operation took place.

As cities get more and more \textit{intelligent}, the concept of smart parking is becoming reality. Ultra low power wireless 
mesh networks with web-based suites of parking management applications are deployed in order to provide real time
parking occupancy status, charging fees and other useful information. A typical smart parking scenario is allowing citizens 
to park without charge at the spaces of their neigborhood. However, such an application could raise important privacy issues as an
attacker could extract from the network private information about the citizens (e.g. Mr. Smith has just arrived home). We propose a
scenario that protects citizen's privacy.      

In this scenario, we assume that there exists a parking space between every two major streets inside a city. Each citizen gets an RFID tag containing 
a private key that encodes his attributes (name, vehicle registration plate and some fresh random value) from the local goverment. The private key 
is used to generate a public key that proves his validity as a citizen. Moreover, depending on his address, a value that encodes the specific 
neighborhood parking space (which should be free only for those who live on one of the two corresponding streets) is added to or substracted from his public key. 
The authority digitally signs the new public key and then places this RFID tag on the citizen's vehicle.

Parking spaces are equipped with RFID tag readers placed on their entrances. Thus, when a citizen wants to park in his neigborhood parking space, 
his device communicates with the reader on the parking entrance and proves two things: first that this vehicle belongs to a valid citizen and 
second that this citizen lives on one of the two neigborhood streets (without revealing which) and is able to park without charge. 
This proof is done by presenting a public key signed from the local goverment (the reader is aware of the trusted authority's public key), by proving that it 
actually holds the private key that generated the citizen validity public key and by proving that the specific parking value has been
added to or substracted from the public key. If the proof is successful the parking entrance bar is lifted and the vehicle is allowed to enter the parking. 
Another citizen's vehicle that is not eligible to park in that space (the parking value has not been added or substracted) is requested to pay 
the appropriate fee in order to enter the parking space and a vehicle that is not registered at all to the local goverment is not allowed enter the parking. 
A malicious user who eavesdrops the proof cannot extract any information that could be useful to him (e.g. in order to get free parking space or 
monitor a citizen's actions). 

\section{Conclusions and Future Work}

In this paper, we considered the problem of implementing zero-knowledge
protocols (ZKP) in low-end devices. Considering the resource limitations of such devices
as well as the restrictions imposed by low power wireless communication protocols 
(e.g. IEEE 802.15.4) we applied the elliptic curve cryptography (ECC) approach. 
Specifically, we have carefully transformed well established zero-knowledge protocols 
based on the discrete logarithm problem (DLP) under the elliptic curve discrete logarithm 
problem (ECDLP). This transformation step was the key for implementing such heavy protocols, 
in terms of computation and communication, on constrained environments, due to the fact that ECC 
offers similar levels of security with other cryptosystems (e.g. RSA) using smaller keys. 
For the first time, we present an implementation of ZKP protocols in an open source and generic 
programming library called Wiselib. Our code is highly portable, freely available and ready to use \cite{wisel}. 
Based on our implementations, we conducted a thorough and comparative evaluation of the protocols on 
two popular hardware platforms equipped with widely used low-end microcontrollers (Jennic JN5139, TI MSP430) 
as well as 802.15.4 RF transceivers. We believe that our results can be used by developers that wish to provide 
certain levels of security and privacy in their applications.  


Future work includes the expansion of the library presented in this
paper so as to include ZKIP protocols which prove various
relations for the encoded values without revealing them (e.g. prove that
my age is over 18 years without revealing it). This way, our library can 
be the basis for implementing attribute based credentials 
\cite{DBLP:conf/ccs/CamenischH02}, \cite{Brands:2000:RPK:517876} on 
embedded devices. 

\bibliographystyle{abbrv}
\bibliography{secdoc}

\appendix
\section{Appendix}

\subsection{Elliptic Curve Cryptography}

For a reader's information, in this section we review some basic concepts of elliptic 
curves and their definition over finite fields. Moreover, we discuss about the
elliptic curve cryptography (ECC) approach and the reasons it consists the best
choice for the implementation of asymmetric cryptography on
constrained devices.

\begin{figure}[H]
  \centering
      \includegraphics[height=1.3in, width=1.6in]{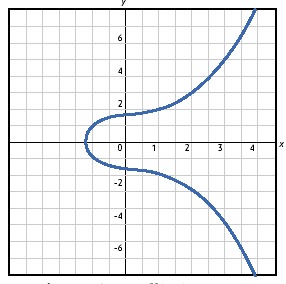}
  \caption{An Elliptic Curve over Real Numbers.}
  \label{fig:el_curve}
\end{figure}

Elliptic curves are usually defined over real numbers (Figure~\ref{fig:el_curve}), 
over a \textit{binary field} $F_{2^m} (m \geq 1)$, or over a \textit{prime field} $F_{p}, (p>3)$. 
An elliptic curve $E$ over a binary field $F_{2^m}$, 
where $m \geq 1$ consists of the set of points $(x,y)$ that satisfy the equation
\begin{equation}
\label{ecc}
y^2 + xy = x^3 + ax^2 +b
\end{equation}
The set of solutions $(x,y)$ of Equation \eqref{ecc} along with a
point \textit{O}, called the point at infinity and a special addition
operation form an elliptic curve group over $F_{2^m}$. The order $m$
of an elliptic curve is the number of points on $E(F_{2^m})$. The
dominant operation in ECC cryptographic schemes is point
multiplication. This operation is the key for the use of
elliptic curves for asymmetric cryptography - the critical operation
which is itself fairly simple, but whose inverse (the elliptic curve
discrete logarithm problem, see below) is computationally hard. The
security of elliptic curve cryptosystems is based on the difficulty
of solving the discrete logarithm problem on the elliptic curve
group. The Elliptic Curve Discrete Logarithm Problem (ECDLP) is
about determining the least positive integer k which satisfies the
equation $Q = k \cdot P$ for two given points $Q$ and $P$ on the
elliptic curve group. The fastest known algorithms for solving the 
ECDLP need exponential time in the worst case, while for solving the same 
problem on non-elliptic curve based groups, the fastest known
algorithms need subexponential time \cite{Koblitz:1998:AAC:275809}. 
For more information on elliptic curves a reader is advised to check 
on the definitive books for ECC, \cite{345194} and \cite{35195}.

From the previous definition, one can realize that Elliptic Curve
Cryptography (ECC), is a relative of discrete logarithm
cryptography. What makes it the best choice for asymmetric
cryptography in comparison with other public key cryptosystmes (e.g.
RSA \cite{DBLP:journals/cacm/RivestSA78} which is based on the problem of factoring large numbers) is
that it offers higher security at the same bits levels. This
advantage exists because of the difference in the method by which a
group is defined, how the elements of the group are defined and how
the fundamental operations are held. These different definitions are
what gives ECC its more rapid increase in security as key length
increases.

\begin{figure}[H]
  \centering
      \includegraphics[height=1.5in, width=2.5in]{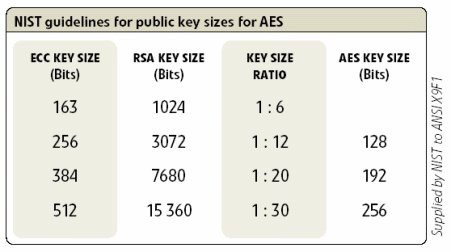}
  \caption{Equivalent Key Sizes and Key Ratio for ECC and RSA.}
  \label{fig:key_equiv}
\end{figure}

Consequently, the reader can realize that the ECC inverse operation
(solving the ECDLP) gets harder, faster, against increasing key
length, in comparison with the inverse operation in RSA (solving the integer
factorization problem). This means that as security
requirements become more stringent, and as processing power gets
more expensive, ECC becomes the more practical system
for use. This keeps ECC implementations smaller and more efficient
than other implementations. As a result, ECC can use a considerably
shorter key and offer the same level of security, whereas other asymmetric
algorithms using much larger ones. Moreover, the gulf between ECC
and its competitors in terms of key size required for a given level
of security becomes dramatically more pronounced, at higher levels
of security \cite{36195}. In Figure~\ref{fig:key_equiv}, one
can see the equivalent key sizes for ECC and RSA.

Conclusively, due to the arguments stated before, ECC is an
excellent choice for doing asymmetric cryptography in portable,
necessarily constrained devices. The smaller ECC keys mean that the
cryptographic operations to be performed by the communicating
devices can be squeezed into considerably smaller hardware, that the
software applications may complete cryptographic operations with
fewer processor cycles and that operations can be performed much
faster, while still guaranteeing equivalent security. In turn, this
means less heat and less power consumption on the devices' chips as
well as software applications that run faster with lower memory
demands.

\end{document}